# Optimum Forward Light Scattering by Spherical and Spheroidal Dielectric Nanoparticles with High Refractive Index

Boris S. Luk`yanchuk*,[1,2], Nikolai V. Voshchinnikov[3], Ramón Paniagua-Domínguez[1], and Arseniy I. Kuznetsov[1]

[1]*Data Storage Institute, A*STAR (Agency for Science, Technology and Research), 5 Engineering Drive 1, 117608, Singapore*
[2]*School of Electrical and Electronic Engineering, Nanyang Technological University, 639798 Singapore*
[3]*Sobolev Astronomical Institute, St Petersburg University, 198504, Russia*

*Corresponding author: Boris_L@dsi.a-star.edu.sg

**ABSTRACT:**

High-refractive index dielectric nanoparticles may exhibit strong directional forward light scattering at visible and near-infrared wavelengths due to interference of simultaneously excited electric and magnetic dipole resonances. For a spherical high-index dielectric, the so-called first Kerker's condition can be realized, at which the backward scattering practically vanishes for some combination of refractive index and particle size. However, Kerker's condition for spherical particles is only possible at the tail of the scattering resonances, when the particle scatters light weakly. Here we demonstrate that significantly higher forward scattering can be realized if spheroidal particles are considered instead. For each value of refractive index $n$ exists an optimum shape of the particle, which produces minimum backscattering efficiency together with maximum forward scattering. This effect is achieved due to the overlapping of magnetic and electric dipole resonances of the spheroidal particle at the resonance frequency. It permits the design of very efficient, low-loss optical nanoantennas.

Resonant nanoparticles and nanoantennas[1] become crucially important for advanced photonic technologies including on-chip interconnects, bioimaging, solar-cells, heat-assisted magnetic recording etc. They can play the role of nanooptical elements, which may substitute conventional optics at subwavelength scale. During the last few years a significant attention has been paid to nanoparticles made of low-loss high-refractive index dielectric and semiconductor materials, in which one can observe both electric and magnetic dipole resonances with comparable strengths at optical frequencies[2-8]. Interference of these two modes allows to fulfill a condition for almost zero backward light scattering, as proposed by Kerker *et al.* for spherical particles more than three decades ago[9,10]. Thus, these materials open a fascinating opportunity to control directionality of scattering and design efficient low-loss nanoantennas[11-13]. The Kerker's-type directional scattering was experimentally demonstrated first for millimeter-scale ceramic spheres in the microwave regime[14], and shortly after for nanometer-scales silicon nanospheres[15] and gallium arsenide nanodisks[16] in the visible spectral range.

In all the above cases, the zero-backward scattering condition has been fulfilled on the long-wavelength tail of the magnetic dipole resonance out of the maximum of the scattering amplitude. However, it was shown recently that for nanostructures with non-spherical shape, namely, flat silicon disks with an aspect ratio around 1:2, electric and magnetic dipole resonances can be overlapped[17], providing a strong forward scattering and almost zero backward scattering at the scattering resonance maximum.

In this paper, we demonstrate that for spheroidal nanoparticles one can always find an optimum aspect ratio, at which the overlapped electric and magnetic dipole resonances provide simultaneously minimal backscattering and optimized forward scattering. This optimum shape depends on the specific value of material refractive index. We work in the frame of exact light scattering methods and consider spheres and spheroids with different aspect ratios.

First, scattering properties of spherical nanoparticles have been analyzed using Mie theory[18] (see Methods for details). Figure 1 presents different scattering characteristics of spherical nanoparticles with radius $R$ versus their refractive index $n$ and size parameter $q = 2\pi R/\lambda$ (here $\lambda$ is the wavelength of incident light). In particular, the partial scattering efficiencies corresponding to $Q_1^{(e)}$ - electric dipole (**ed**), $Q_1^{(m)}$ - magnetic dipole (**md**), $Q_2^{(e)}$ - electric quadrupole (**eq**) and $Q_2^{(m)}$ - magnetic quadrupole (**mq**) are shown in Fig.1a for a sphere with $n = 2.4$. The total scattering efficiency can be accurately described in this range of $q$ values as a sum of these four partial resonant efficiencies. The two observed dominant peaks can be identified with the resonant excitation of the magnetic dipolar and magnetic quadrupolar modes. We can also see pronounced minima in the backscattering cross-sections depicted in Fig. 1b. These minima correspond to the so called "first Kerker's condition" $a_1 = b_1$, for which the amplitudes and phases of electric and magnetic dipole resonances are equal. This condition is marked by filled circles in Fig. 1c. Satisfying the condition $a_1 = b_1$ for both real and imaginary part leads to the solution $nq = \alpha = const$, where $\alpha \approx 2.7437$ is the root of the equation:

$$1 + (2\alpha^2 - 1)\cos(2\alpha) + \alpha(\alpha^2 - 2)\sin(2\alpha) = 0.$$

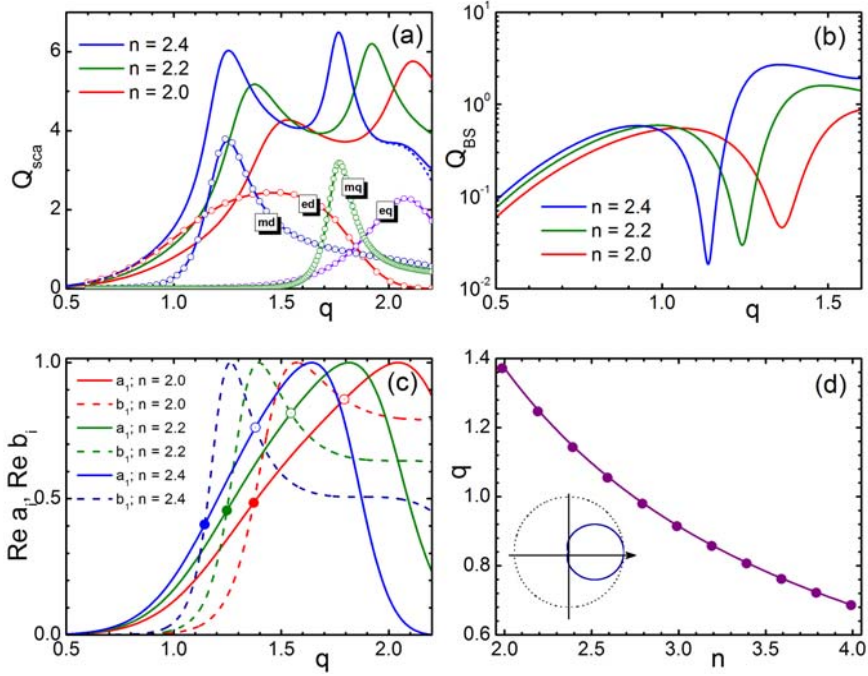

**Figure 1.** (a) Total scattering efficiency, $Q_{sca}$, versus size parameter $q = 2\pi R/\lambda$ for three different values of refractive index $n = 2.0$, 2.2 and 2.4. Four partial scattering efficiencies (curves with open circles) are shown for $n = 2.4$ corresponding to the electric dipole (ed), magnetic dipole (md), electric quadrupole (eq) and magnetic quadrupole (mq) contributions. (b) Backscattering efficiency $Q_{BS}$ (in logarithmic scale) versus size parameter $q$ for three different values of refractive index $n = 2.0$, 2.2 and 2.4, exhibiting a pronounced minima at particular values of size parameter. (c) Values of electric dipole coefficient $a_1$ (solid lines) and magnetic dipole coefficient $b_1$ (dashed lines). Positions in which $\mathrm{Re}\, a_1 = \mathrm{Re}\, b_1$ and $\mathrm{Im}\, a_1 = \mathrm{Im}\, b_1$, corresponding to the first Kerker condition[9,10], are plotted as filled circles. The associated values of $q$ correspond with high accuracy to the minimum values of $Q_{BS}$. For open circles one have $\mathrm{Re}\, a_1 = \mathrm{Re}\, b_1$ but $\mathrm{Im}\, a_1 = -\mathrm{Im}\, b_1$. (d) Trajectory of minimum back scattering on the plane of parameters $\{n, q\}$. Solid line presents solution to the equation $a_1 = b_1$ while circles are numerical solutions to the equation $Q_{BS} \to \min$. Inset in plot (d) shows polar scattering diagram[19] along the trajectory $Q_{BS} \to \min$ with pronounced forward scattering.

In Fig. 1d we present the trajectory of minimum backscattering on a $q,n$ parameters' plane. From this figure one can conclude that minimum backscattering is well described by solution of $nq = \alpha$ not only for $q \ll 1$ (i.e. conventional condition for applicability of the dipole approximation) but even for values of size parameter $q$ of the order of unity. The reason for this effect can be seen in Fig. 1a for $n = 2.4$ and for a general case in Supplementary Figure 1S. Along the trajectory $q = \alpha/n$ and for a refractive index above 1.5 higher order modes (quadrupoles, etc.) are strongly suppressed compared to the dominant dipole modes.

In Fig. 2, we show a contour plot of the backscattering cross-section where one can see the trajectory of the pronounced minimum of $Q_{BS}(n,q)$ (dashed curve 3). Trajectory of the maximum value of $Q_{FS}/Q_{BS}$ (circles) practically coincides with curve 3. Additionally, trajectories of the maximum value of total scattering $Q_{sca}(n,q)$ (curve 1) and maximum forward scattering $Q_{FS}(n,q)$ (curve 2) are also shown in the figure. All these curves follow approximately hyperbolic dependence on the refractive index and, consequently, do not cross each other. This means that for a spherical particle, whatever the particle parameters are, it is not possible to obtain resonant values of the total $Q_{sca}(n,q)$ or forward $Q_{FS}(n,q)$ scattering efficiencies along the trajectory fulfilling the first Kerker's condition for the minimum backward scattering $a_1 = b_1$. From formula (2), see Methods, the latter condition leads to $Q_{FS} \propto |a_1|^2/q^2 \propto n^2|a_1|^2$ and therefore the maximum value of $Q_{FS}$ corresponds to the maximum value of $n^2|a_1|^2$ (see inset in Fig. 2). However, the values of scattering amplitudes $|a_1| = |b_1|$ at maximum $Q_{FS}$ are quite small, below 0.5 (see Fig. 1c). It is clear that if we would be able to overlap electric and magnetic resonances at the point $a_1 = b_1 \approx 1$ we could enhance both total and forward scattering values.

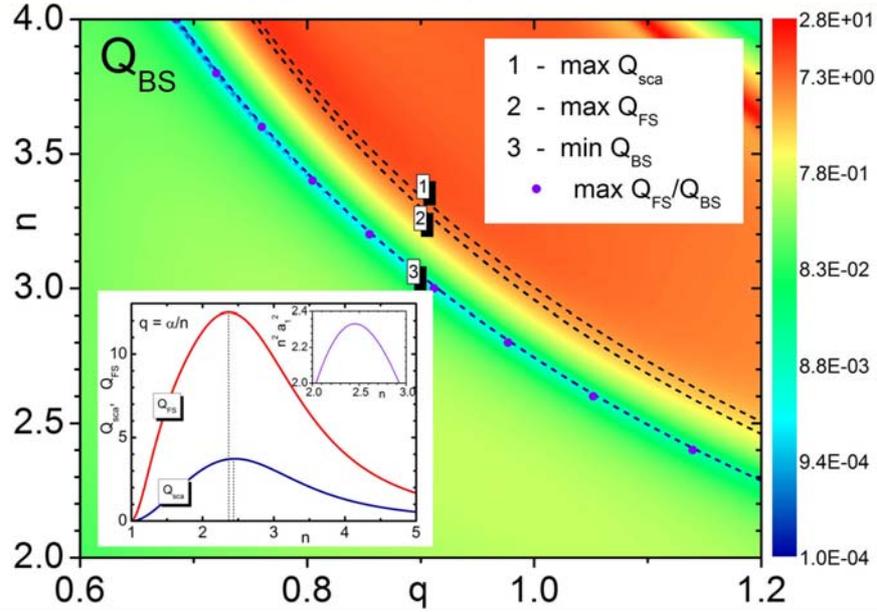

**Figure 2.** Contour plot of the backscattering efficiency $Q_{BS}(q,n)$ on the $q, n$ parameters' plane. Dashed lines show trajectories of maximum total $Q_{sca}$ (curve 1) and forward $Q_{FS}$ (curve 2) scattering efficiencies. Curve 3 shows the trajectory of the minimum back scattering efficiency. Inset shows the variation of $Q_{sca}$ and $Q_{FS}$ along the trajectory $q = q(n)$, where $Q_{BS}$ reaches minimum (i.e., along curve 3). The upper right part of the inset presents the function $|a_1(n)|^2 n^2$ along the trajectory of the first Kerker's condition.

One of the possibilities to satisfy condition $a_1 = b_1 \approx 1$ is to use metallic-dielectric core-shell nanoparticles[20]. It can also be reached by changing the particle's shape, e.g. using oblate spheroidal nanoparticles instead of spheres. As it was shown in our previous work (see Fig. 4 in Ref. 15), squeezing a silicon sphere into a spheroid with aspect ratio around 1:2 it is possible to obtain overlapping between the electric and magnetic dipole resonances and minimized backward scattering close to the wavelength of scattering resonances. This is also consistent with results published later for silicon nanodisks[17]. Going one step further, we will now demonstrate that for any given value of the particle refractive index there is a particular particle shape at which a resonant forward scattering with minimized backward scattering can be realized.

Solution of the wave equation in spheroidal coordinates can be made using the separation of variables method[21] (see also Methods). In the following we will focus our study on oblate spheroids, since for this shape electric and magnetic dipole resonances can be overlapped. An oblate spheroid (ellipsoid of revolution) is obtained by the rotation of an ellipse with focal distance $d$ around its minor axis. The ratio of the major semiaxis $a$ to the minor semiaxis $b$ (i.e. the aspect ratio $a/b$) characterizes the particle shape which may vary from a nearly spherical ($a/b \approx 1$) to a disk one ($a/b \gg 1$). The particle size can be specified by parameter $g = \frac{2\pi}{\lambda}\frac{d}{2}$ related to the length of the focal distance or by parameter $q_v = \frac{2\pi r_v}{\lambda}$, related to the radius of the sphere $r_v$ whose volume is equal to that of the spheroid. For oblate spheroids, $r_v^3 = a^2 b$. The connection between $g$ and $q_v$ is given by

$$q_v = g \frac{(a/b)^{2/3}}{\left[(a/b)^2 - 1\right]^{1/2}} = \frac{2\pi a}{\lambda}(a/b)^{-1/3},$$

where the size parameter $q = \frac{2\pi a}{\lambda}$ plays the same role as parameter $\frac{2\pi R}{\lambda}$ in the Mie theory.

In general, the angle $\beta$ between the propagation direction and the rotation axis of the spheroid can be arbitrary ($0^o \leq \beta \leq 90^o$). Here, we study the case $\beta = 0$ when radiation propagates along the minor axis.

In Fig. 3 we show the total $\tilde{Q}_{sca}$ and forward $\tilde{Q}_{FS}$ scattering efficiencies (see Methods for details) of oblate spheroidal particles with different refractive index $n$ versus aspect ratio $a/b$ along the trajectories of minimum backward scattering. It can be seen from the figure that for each value of the particle refractive index there is an aspect ratio of the spheroid $a/b$ for which the scattering efficiencies are optimized. As a general rule, higher values of scattering efficiencies can be achieved with higher refractive indices. This is not the case of spherical particles whose directional scattering is optimized at refractive index of ~2.45 (see inset in Fig.2). For each particular material one can find an optimum spheroidal shape which will produce a maximum forward scattering at a minimum back scattering.

One can compare the efficiency of forward scattering by spherical and spheroidal particles at the minimized backscattering condition. For example, spherical particle with $n \approx 2.4$ has maximum forward scattering $Q_{FS} = \tilde{Q}_{FS} \approx 12.3$ (see insert in Fig. 2). An optimized spheroidal particle with the same refractive index and $a/b \approx 1.7$ has about 1.4 times higher forward scattering efficiency. With higher refractive index this difference becomes much more pronounced. For spherical particle with $n \approx 4$ one can reach maximum $\tilde{Q}_{FS} \approx 4$. An oblate particle with $n \approx 4$ and $a/b \approx 2.25$ permits to reach $\tilde{Q}_{FS}$ value above 30 with the back scattering close to zero. *This presents a huge interest for optical antennas.* Naturally, deformation of the particle shape leads also to some shift in the position of the resonant frequency.

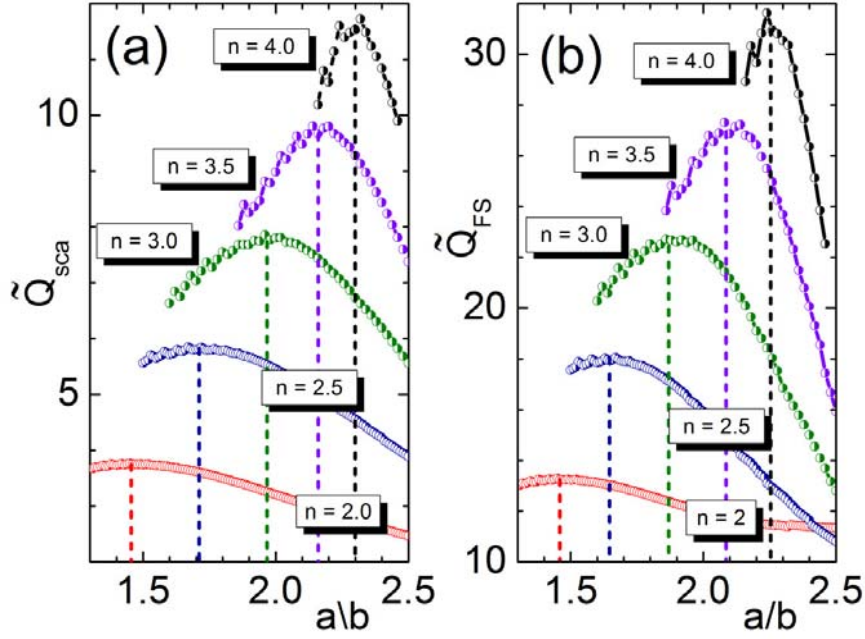

**Figure 3.** Variation of total $\tilde{Q}_{sca}$ (a) and forward $\tilde{Q}_{FS}$ (b) scattering efficiencies along the trajectories of minimum backward scattering for oblate spheroidal particles with different refractive index $n$ versus aspect ratio $a/b$.

Final result of optimization is presented in Fig. 4. It shows the dependence of the optimum shape of the spheroids on the particle refractive index: $a/b = f(n)$, which corresponds to maximum of $\tilde{Q}_{sca}$ or $\tilde{Q}_{FS}$ (solid lines) with minimized $\tilde{Q}_{BS}$. Resonant frequencies follow from the resonant values of the size parameter $q = q(n)$ related to these shapes (shown by corresponding dashed curves). For example, for $n = 3.5$ one can find from Fig. 4 values $a/b \approx 2.09$ and $q \approx 1.28$ for optimum forward scattering.

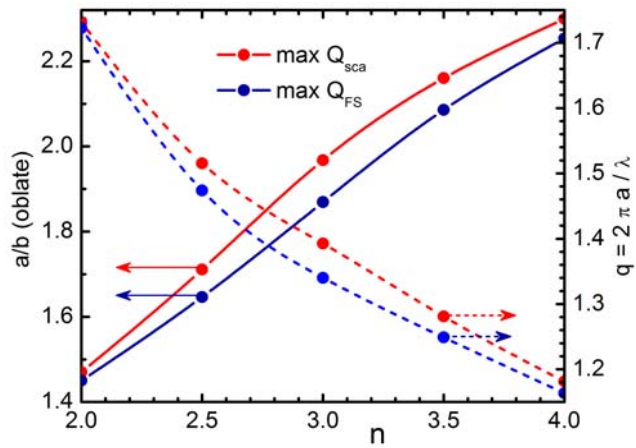

**Figure 4.** Optimum shape $a/b$ (solid lines) and size parameter $q$ (dashed lines) for spheroidal particles versus value of refractive index $n$.

As we mentioned above, the physical reason for scattering maximization is related to overlapping of magnetic and electric dipole resonances of the particles. Dynamics of this overlapping for particles with refractive index $n = 3.5$ is shown in Fig. 5. For spherical particle $a/b = 1$ magnetic (md) and electric (ed) dipole resonances are well separated. It can be easily seen from the corresponding partial scattering efficiencies calculated from the Mie theory. Similar multipole decomposition for partial efficiencies can be done for spheroidal particles as well (see Methods for details). According to Fig. 4 optimum condition for forward scattering is reached for $a/b = 2.086$. In Fig. 5 one can see that with increase of the aspect ratio the electric and magnetic dipole resonances approach each other and fully merged at $a/b = 2.086$, which allows obtaining minimum backward scattering condition at the resonance of total scattering.

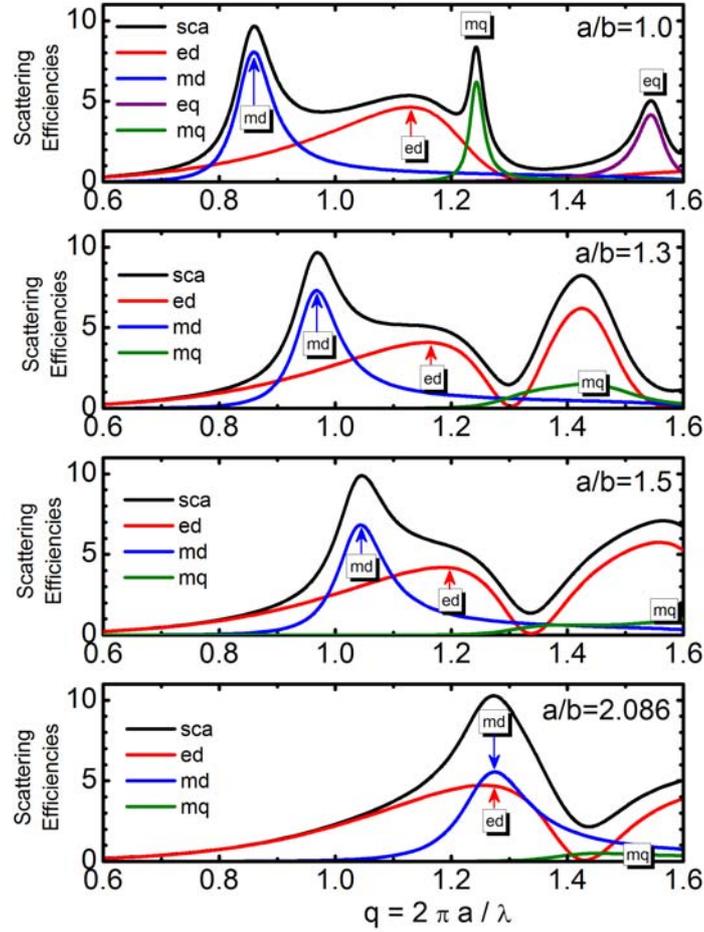

**Figure 5.** Overlapping of electric and magnetic dipole resonances for different particle shapes. Total scattering efficiency (black line), together with the corresponding electric $a_1$ (red line) and magnetic $b_1$ (blue line) scattering dipolar contributions are plotted versus size parameter $q$ for different values of the aspect ratio $a/b$, ranging from 1 (sphere) to the optimized value 2.086. Green curve plots the scattering contribution of the magnetic quadrupole mode.

In conclusion, we have investigated the problem of the shape optimization for oblate spheroidal dielectric particles aimed to obtain minimum backscattering together with maximized total and forward scattering. We have shown that this optimization is possible for any given value of refractive index. Such optimized particles are extremely efficient directional optical nanoantennas, which can act as Huygens sources[11]. Efficiency of spheroidal scatterers depends on the refractive index of the material and can be significantly higher than those which can be reached with spherical particles.

**Acknowledgements.** This work was supported by Data Storage Institute (DSI) core funds.

**Methods**

*Scattering by spheres*

Scattering efficiencies for total, $Q_{sca}$, forward, $Q_{FS}$, and backward, $Q_{BS}$, scattering of a spherical particle can be derived within the framework of Mie theory[18] and written as:

$$Q_{sca} = \frac{2}{q^2} \sum_{\ell=1}^{\infty} (2\ell+1)\left(|a_\ell|^2 + |b_\ell|^2\right), \tag{1}$$

$$Q_{FS} = \frac{1}{q^2} \left| \sum_{\ell=1}^{\infty} (2\ell+1)(a_\ell + b_\ell) \right|^2, \tag{2}$$

$$Q_{BS} = \frac{1}{q^2} \left| \sum_{\ell=1}^{\infty} (2\ell+1)(-1)^\ell (a_\ell - b_\ell) \right|^2. \tag{3}$$

The electric, $a_\ell$, and magnetic, $b_\ell$, scattering amplitudes for nonmagnetic materials with relative magnetic susceptibility $\mu = 1$, and dielectric permittivity $\varepsilon = n^2$ ($n$ being the refractive index of the particle material) are given by:

$$a_\ell = \frac{\Re_\ell^{(a)}}{\Re_\ell^{(a)} + i\, \Im_\ell^{(a)}}, \qquad b_\ell = \frac{\Re_\ell^{(b)}}{\Re_\ell^{(b)} + i\, \Im_\ell^{(b)}}, \tag{4}$$

where $\Re_\ell$ and $\Im_\ell$ functions are defined as follows:

$$\Re_\ell^{(a)} = n\psi_\ell'(q)\psi_\ell(nq) - \psi_\ell(q)\psi_\ell'(nq), \quad \Im_\ell^{(a)} = n\chi_\ell'(q)\psi_\ell(nq) - \chi_\ell(q)\psi_\ell'(nq), \tag{5}$$

$$\Re_\ell^{(b)} = n\psi_\ell'(nq)\psi_\ell(q) - \psi_\ell(nq)\psi_\ell'(q), \quad \Im_\ell^{(b)} = n\chi_\ell'(q)\psi_\ell(nq) - \psi_\ell(nq)\chi_\ell'(q). \tag{6}$$

Here, $\psi_\ell(z) = \sqrt{\dfrac{\pi z}{2}} J_{\ell+\frac{1}{2}}(z)$, $\chi_\ell(z) = \sqrt{\dfrac{\pi z}{2}} N_{\ell+\frac{1}{2}}(z)$, where $J_{\ell+\frac{1}{2}}(z)$ and $N_{\ell+\frac{1}{2}}(z)$ are the Bessel and Neumann functions. The radius of the particle $R$ enters in this theory through the dimensionless size parameter $q = \omega R/c = 2\pi R/\lambda$, where $\omega$ is the angular frequency, $c$ the speed of light, and $\lambda$ the radiation wavelength in vacuum. The prime in formulas (5), (6) indicates differentiation with respect to the argument of the function, i.e. $\psi'_\ell(z) \equiv d\psi_\ell(z)/dz$, etc. The efficiencies (1)-(3) represent the corresponding cross sections normalized to the geometrical cross section of the sphere. The total scattering efficiency is then given by sum of partial scattering efficiencies:

$$Q_{sca} = \sum_{\ell=1}^{\infty} \left( Q_\ell^{(e)} + Q_\ell^{(m)} \right), \qquad (7)$$

where each partial efficiency corresponds to the radiation of the $\ell$-th order multipole. Terms $Q_\ell^{(e)}$ and $Q_\ell^{(m)}$ describe the radiation related to the electric and magnetic polarizabilities, respectively. In the following, we will discuss transparent dielectrics with $\mathrm{Im}\,\varepsilon = 0$, so $Q_{ext} = Q_{sca}$.

*Scattering by spheroids*

The optical properties of spheroidal particles can be determined by various methods of light scattering theory. Most frequently, the separation of variables method and the T-matrix method are used. The survey of methods can be found in Ref. 22 (see also Database of Optical Properties of cosmic dust analogues, DOP, http://www.astro.spbu.ru/DOP/3-REVS/index.html).

Asano & Yamamoto[23] obtained the first solution to the light scattering problem for spheroids with a complex refractive index. The method is based on the solution to the Helmholtz equation in the spheroidal coordinate system. Asano & Yamamoto applied the Debye potentials to describe the electromagnetic fields, which is similar to the Mie solution for spheres. The scattering coefficients then are found in the infinite systems of the linear algebraic equations and can be found by solving truncated systems.

Another solution was published by Farafonov[24] (see Ref. 25 for first numerical results). Its principal distinction from the previous one is the special basis for the representation of the electromagnetic fields - a combination of the Debye and Hertz potentials (i.e. the potentials introduced to solve the light scattering problem for spheres and infinitely long cylinders, respectively). The approach has an incontestable advantage for strongly elongated or flattened particles.

In this paper, we use the most recent version of the numerical code based on the Farafonov's solution (see http://www.astro.spbu.ru/DOP/6-SOFT/SPHEROID/1-SPH_new/). The comparison of methods and benchmark results can be found in Ref. 26.

For spheroids, one usually calculates the scattering efficiency factors $Q = \sigma/S$ which are the ratios of the corresponding cross-sections $\sigma$ to the geometrical cross-section $S$ of the spheroid (the area of the particle's shadow). For oblate spheroids and $\beta = 0$, $S = \pi a^2$. The efficiencies for forward, $Q_{FS}$, and backward, $Q_{BS}$, scattering are[27] (pay attention that Eq. (5.80) for backscattering efficiency in Ref. 27, p.137 must be corrected):

$$Q_{FS} = \frac{1}{q_v^2 (a/b)^{2/3}} \left| \sum_{\ell=1}^{\infty} i^{-\ell} b_\ell^{(1)} \sum_{r=0,1}^{\infty}{}' d_r^{1\ell}(-ig)(r+1)(r+2) \right|^2, \quad (8)$$

$$Q_{BS} = \frac{1}{q_v^2 (a/b)^{2/3}} \left| \sum_{\ell=1}^{\infty} i^{\ell} b_\ell^{(1)} \sum_{r=0,1}^{\infty}{}' d_r^{1\ell}(-ig)(r+1)(r+2) \right|^2. \quad (9)$$

Here, $b_\ell^{(1)}$ are the coefficients for scattered radiation which are determined from the solution to the light scattering problem and $d_r^{1\ell}(-ig)$ are the expansion coefficients of oblate angular spheroidal functions in terms of associated Legendre polynomials. The prime over the summation symbols indicates the even (odd) terms only are summarized when the index $(\ell - 1)$ is even (odd).

A convenient way to compare the optical properties of particles with different shapes is normalizing the cross-sections by the geometrical cross-sections of the equal volume spheres, $\sigma/\pi r_v^2$. For oblate spheroids and $\beta = 0^o$, we have

$$\tilde{Q} = \frac{\sigma}{\pi r_v^2} = (a/b)^{2/3} Q. \quad (10)$$

*Multipole decomposition*

Multipole decomposition allows one to identify the multipolar character of the different resonances being excited in a system[28-30]. In our case, it was performed by projecting the electromagnetic field scattered by the spheroids into the Vector Spherical Harmonics basis on a spherical surface with radius $R_0$ enclosing the structure. The center of the sphere was chosen to coincide with the center of the spheroid. In this way, one can compute the electric $a_{lm}$ and magnetic $b_{lm}$ scattering coefficients associated to a certain multipolar contribution as:

$$a_{lm} = \frac{(-i)^{l+1} kR_0}{h_l^{(1)}(kR_0)[\pi(2l+1)l(l+1)]^{1/2} E_0} \iint Y_{lm}^*(\theta,\phi)\hat{\mathbf{r}} \cdot \mathbf{E}_{sca} d\Omega, \qquad (11)$$

$$b_{lm} = \frac{\eta(-i)^{l} kR_0}{h_l^{(1)}(kR_0)[\pi(2l+1)l(l+1)]^{1/2} E_0} \iint Y_{lm}^*(\theta,\phi)\hat{\mathbf{r}} \cdot \mathbf{H}_{sca} d\Omega, \qquad (12)$$

where $h_l^{(1)}(kR_0)$ is the spherical Hankel function of first kind and order $l$ and:

$$Y_{lm}(\theta,\phi) = \sqrt{\frac{2l+1}{4\pi}\frac{(l-m)!}{(l+m)!}} P_l^m(\cos\theta) e^{im\phi}, \qquad (13)$$

with $P_l^m(\cos\theta)$ being the associated Legendre polynomials. The partial scattering efficiency due to the $l$-th electric or magnetic multipole can then be computed as:

$$Q_l^E = \frac{2l+1}{q^2} \sum_{m=-l}^{l} |a_{lm}|^2, \qquad (14)$$

$$Q_l^M = \frac{2l+1}{q^2} \sum_{m=-l}^{l} |b_{lm}|^2. \qquad (15)$$

$Q$ or $\tilde{Q}$ is obtained depending on whether one uses $q$ or $q_v$. The total scattering efficiency can be retrieved by summing up the contributions of the different electric and magnetic multipoles. While the choice of the radius of the sphere is arbitrary, the only requirement to achieve accurate results is a sufficiently accurate angular resolution in the integral.